\documentclass[%
 reprint,
superscriptaddress,
 amsmath,amssymb,
 aps,
prl,
]{revtex4-1}

\usepackage{graphicx}% Include figure files
\usepackage{dcolumn}% Align table columns on decimal point
\usepackage{bm}% bold math
\usepackage{bbold}
\begin{document}

\preprint{APS/123-QED}

\title{Generative embeddings of brain collective dynamics using variational autoencoders}

%\title{Variational Autoencoders generate\\the undelying structural connectivity and states of consciousness modeling human sleep cycle}% Force line breaks with \\

\author{Yonatan Sanz Perl}
  \affiliation{Universidad de San Andrés}%
  \affiliation{Physics Department (University of Buenos Aires) and Buenos Aires Physics Institute (CONICET)}%
 \author{Hernán Boccacio}%
\affiliation{Physics Department (University of Buenos Aires) and Buenos Aires Physics Institute (CONICET)}%
\author{Ignacio Pérez-Ipiña}%
\affiliation{Physics Department (University of Buenos Aires) and Buenos Aires Physics Institute (CONICET)}%
 \author{Federico Zamberlán}%
\affiliation{Physics Department (University of Buenos Aires) and Buenos Aires Physics Institute (CONICET)}%
  \author{Helmut Laufs}%
    \affiliation{Department of Neurology, Christian-Albrechts-University Kiel}
  \author{Morten Kringelbach}%
  \affiliation{Department of Psychiatry, University of Oxford}%
  \author{Gustavo Deco}%
 \affiliation{Center for Brain and Cognition, Computational Neuroscience Group, Universitat Pompeu Fabra}%
 \author{Enzo Tagliazucchi}%
\affiliation{Physics Department (University of Buenos Aires) and Buenos Aires Physics Institute (CONICET)}%

\date{\today}% It is always \today, today,

\begin{abstract}

We consider the problem of encoding pairwise correlations between coupled dynamical systems in a low-dimensional latent space based on few distinct observations. We used variational autoencoders (VAE) to embed temporal correlations between coupled nonlinear oscillators that model brain states in the wake-sleep cycle into a two-dimensional manifold. Training a VAE with samples generated using two different parameter combinations resulted in an embedding that represented the whole repertoire of collective dynamics, as well as the topology of the underlying connectivity network. We first followed this approach to infer the trajectory of brain states measured from wakefulness to deep sleep from the two endpoints of this trajectory; next, we showed that the same architecture was capable of representing the pairwise correlations of generic Landau-Stuart oscillators coupled by complex network topology

%How a dynamical system can be learned by deep learning generative models and how this cross fertilisation can be fruitful in other disciplines is an outstanding open question in science. We approach this question by training a variational autoencoder (VAE) using simulated functional connectivity matrices yielded by whole brain dynamical models that fits empirical data from human deep sleep and wake state . We found that the VAE is capable to generate functional connectivity from other sleep states that not were included in the training set, and also is capable to generate the underlying structural connectivity. We show the relation existing between the latent variables and the model parameters in the homogeneous whole brain model. 

\end{abstract}

\maketitle

%\tableofcontents

Several  biological systems can be understood in terms of simple dynamical rules coupled by heterogeneous connectivity patterns. Perhaps the most paradigmatic case is the human brain, where complex collective behaviour emerges from the nonlinear dynamics of $\approx 10^{10}$ neurons interacting at $\approx 10^{15}$ synaptic connections \cite{chialvo2010emergent}. In spite of this complexity at the microscopic scale, the brain spontaneously self-organises into a reduced number of discrete states, such as those in the wake-sleep cycle, which suggests that a low-dimensional manifold is sufficient to encode its large-scale dynamics \cite{cavanna2018dynamic}. 

The mechanisms underlying the emergence of different brain states can be probed using whole-brain models based on conceptually simple local dynamical rules coupled according to empirical measurements of anatomical connectivity \cite{deco2012ongoing,breakspear2017dynamic,haimovici2013brain}, for instance, by coupling nonlinear oscillators with the long-range white matter tracts inferred from diffusion tensor imaging (DTI) \cite{deco2017dynamics}. After parameter optimisation to reproduce neuroimaging data acquired during different brain states, the models can be used to explore the interplay between local dynamics, long-range structural coupling, and the formation of large-scale activity patterns \cite{ipina2020modeling,deco2018whole,kringelbach2020dynamic}, and as methods for data augmentation to be combined with machine learning techniques for the purpose of brain state classification \cite{perl2020data,arbabyazd2020completion}.

While whole-brain models can reproduce the functional connectivity of brain states such as those seen in the progression from wakefulness to deep sleep \cite{jobst2017increased, ipina2020modeling}, it is unclear whether coupled dynamical systems can also capture relationships between these states, encoding them into a low dimensional manifold that preserves the ordering within progressions of brain states. More generally, we consider a system of coupled units whose dynamics have been optimised to reproduce the second-order statistics (i.e. pairwise correlations) of a real-world system, and ask whether different discrete states of such system can be efficiently represented by latent variables that are capable of reproducing the whole repertoire of states from a reduced number of representative examples. In the particular case of collective brain dynamics, this is equivalent to asking whether the endpoint states of a certain progression, such as the descent from wakefulness into deep sleep, can be used to learn a latent representation which encodes all intermediate stages, and can be extrapolated to produce correlations corresponding to states beyond this progression.

We used whole-brain models fitted to empirical data to generate pairwise correlation matrices for the different brain states that comprise the human wake-sleep cycle: wakefulness, N1, N2 and N3 sleep (N1 and N2 are intermediate stages, while N3 is the deepest stage of human sleep) \cite{berry2012rules}. Next, we trained a variational autoencoder (VAE) with matrices corresponding to wakefulness and N3 sleep, showing that intermediate (N1 and N2) sleep stages were embedded continuously in the latent space, and that the resulting two-dimensional manifold also extrapolated to capture known results concerning the structure-function relationship during unconsciousness \cite{vincent2007intrinsic, barttfeld2015signature, tagliazucchi2016deep}. Finally, we assessed the relationship between latent space variables and the parameters of generic coupled Stuart-Landau oscillators.

\emph{Whole-brain model.—} We start from a model constructed from 90 Stuart-Landau nonlinear oscillators, each representing the dynamics within a macroscopic brain region of interest \cite{deco2017dynamics}. The coupled dynamics are given by, 

\begin{eqnarray}
\label{eq:one}
\frac{dx_j}{dt}=(a-x_j^2-y_j^2)x_j-\omega_jy_j+ \nonumber \\ G\sum_{i}J_{ij}(x_i-x_j)+\beta
\eta_j \\
\frac{dy_j}{dt}=(a-x_j^2-y_j^2)y_j+\omega_jx_j+ \nonumber \\ G\sum_{i} J_{ij}(y_i-y_j)+\beta\eta_j \nonumber
\end{eqnarray}

Where $x_j$ is the dynamical variable that simulates the functional magnetic resonance (fMRI) signal of region $j$ and $J_{ij}$ represents the symmetrical coupling matrix that weights the connectivity between regions $i$ and $j$. This matrix is inferred from the diffusion of water molecules in white matter from DTI recordings, and represents the empirical distribution of long-range axon bundles in the brain. The bifurcation parameter ($a$) controls the proximity to oscillatory dynamics and $G$ globally scales the coupling between oscillators. Finally, $\eta_j$ is an additive Gaussian noise term that is scaled by $\beta=0.04$. 

Eq. \ref{eq:one} can be optimised to reproduce the second-order statistics of fMRI data acquired during different brain states. In previous work, we proposed to reduce the complexity of the model by grouping brain regions into well-studied functional networks, known as resting state networks (RSN) \cite{ipina2020modeling}. We encoded the 90 bifurcation parameters ($a_j$) into 6 parameters representing the contribution of each RSN to the local dynamics as $a_j = \sum_k^6\Delta_k \mathbb{1}_{jk}$, where $\mathbb{1}_{jk}$ equals 1 if the node $j$ belongs to the $k$-th RSN and zero otherwise. We then applied a stochastic optimisation algorithm to determine the $\Delta_k$ and $G$ that best reproduce the correlation matrix $C_{ij}$ of each state in the progression from wakefulness to deep sleep. The $C_{ij}$ contains in its $i,j$ entry the linear correlation between the empirical/simulated fMRI time series corresponding to nodes $i$ and $j$ \cite{ipina2020modeling}.

\begin{figure}
\includegraphics[scale=0.25]{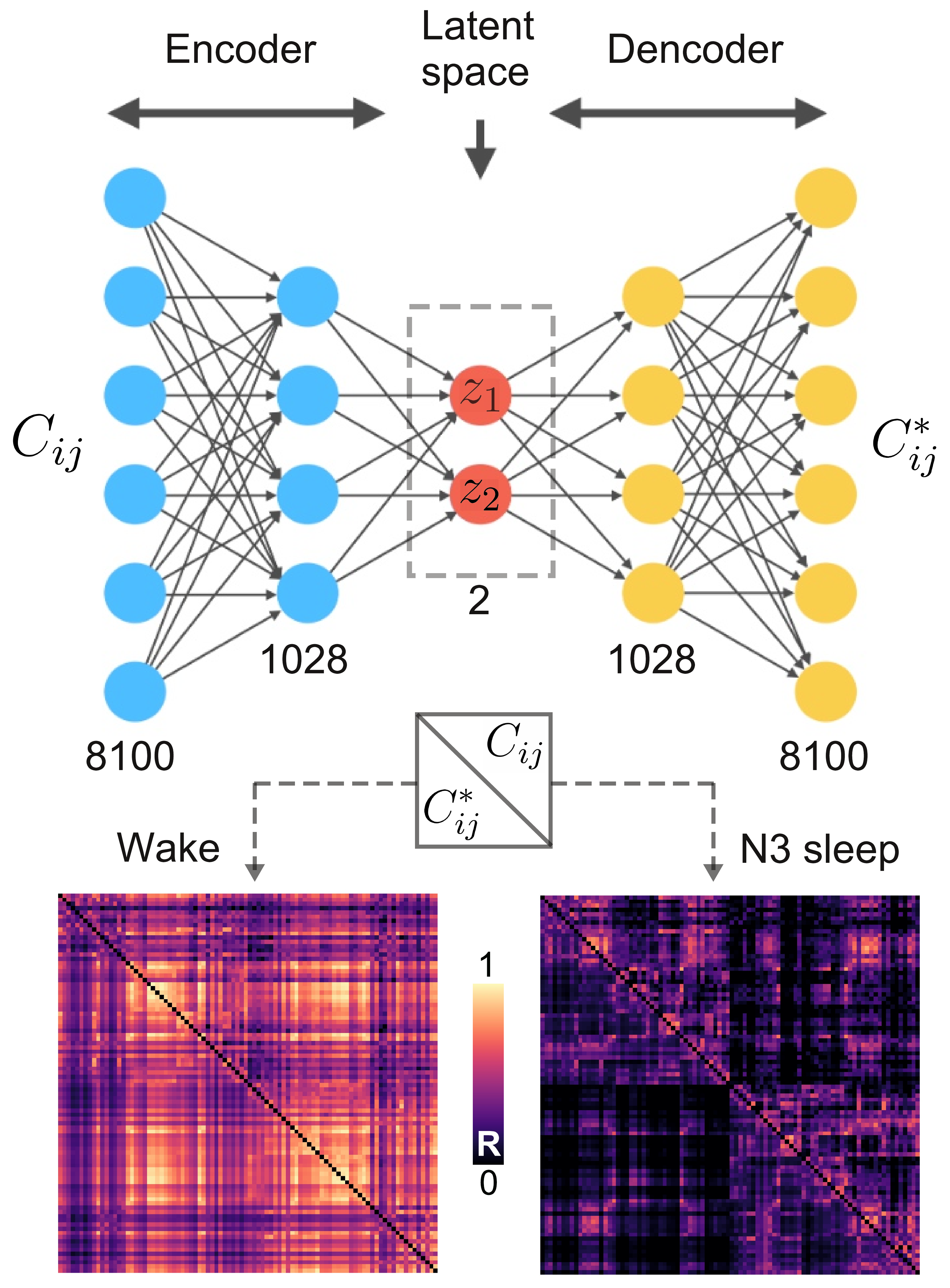}%Here is how to import EPS art
\caption{\label{scheme} VAE architecture. The inputs are correlation matrices $C_{ij}$ obtained from the model (Eq. \ref{eq:one}) fitted to wakefulness and N3 sleep. The input layer has 8100 units, followed by an intermediate layer with 1028 neurons and a two-dimensional latent space. The next two layers reverse the encoding process, yielding a matrix $C_{ij}^*$ for each $z_1$, $z_2$ pair in the latent space. The bottom panel presents input matrices $C_{ij}$ (above diagonal) and their reconstructed versions $C_{ij}$ (below diagonal) for the model fitted to wakefulness and N3 sleep.}
\end{figure}

\emph{Encoding the $C_{ij}$ with a VAE.—}
We implemented a VAE to find a low-dimensional representation encoding the progression of brain states. VAE are autoencoders trained to map inputs to probability distributions in latent space, which can be regularised during the training process to produce meaningful outputs after the decoding step. The architecture of the implemented VAE (shown in Fig.~\ref{scheme}) can be subdivided into three parts: the encoder network, the middle variational layer, and the decoder network. The encoder consists of a deep neural network with rectified linear units (ReLu) as activation functions and two dense layers, which bottlenecks into the two-dimensional variational layer, where units $z_1$ and $z_2$ span the latent space. The encoder network applies a nonlinear transformation to map the $C_{ij}$ into Gaussian probability distributions in latent space, and the decoder network mirrors the encoder architecture to produce reconstructed matrices $C_{ij}^*$ from samples of these distributions. 

To train the network, the errors were backpropagated via gradient descent with the purpose of minimising a loss function composed of two terms: a standard reconstruction error term (computed from the units in the output layer of the decoder), and a regularisation term computed as the Kullback-Leibler divergence between the distribution in latent space and a standard Gaussian distribution. The regularisation term ensures continuity and completeness in the latent space, i.e. that similar values are decoded into similar outputs, and that those outputs represent meaningful combinations of the encoded inputs. \cite{kingma2013auto}. 

We generated 5000 correlation matrices $C_{ij}$ corresponding to wakefulness and N3 sleep using the model described in Eq. \ref{eq:one}. We then created 80\%/20\% random splits to obtain training and test sets, and used the training set to optimise the VAE parameters. The training procedure consisted of batches with 128 samples and 50 training epochs using an Adam optimiser and the loss function described in the previous paragraph.
%\begin{figure*} con esto la pone ocupando las dos columnas

\begin{figure}
\includegraphics[scale=0.29]{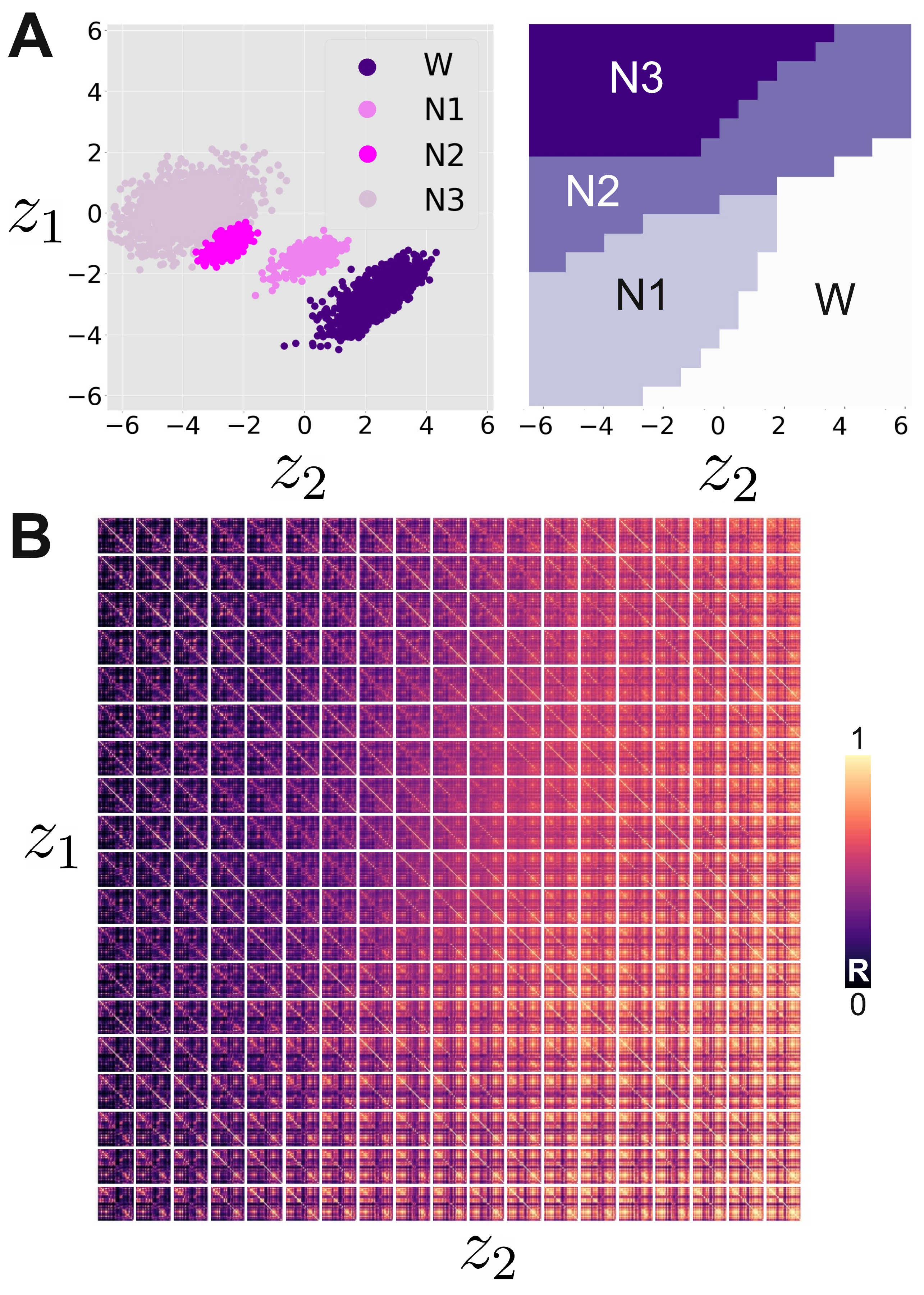}
\caption{\label{multi} The latent space obtained from wakefulness and N3 sleep contains the orderly progression of intermediate brain states. A) Left: Latent state representation obtained by encoding the test set (wakefulness and N3 sleep), and the encoding obtained for the two intermediate states that were not used to train the VAE (N1 and N2 sleep). Right: Latent space divided into regions with maximal similarity to wakefulness, N1, N3 and N3 correlation matrices. B) Correlation matrices obtained by decoding an exhaustive exploration of the latent space variables $z_1$ and $z_2$.}
\end{figure}

\emph{The latent space encodes the progression of brain states during sleep.—} The encoding process applied to the wakefulness and N3 sleep data generated two distinct clusters in the latent space (Fig.~\ref{multi}A, left). The encoding of the correlation matrices corresponding to intermediate sleep stages not used to train the VAE (N1 and N2 sleep) resulted in separate clusters organised according to sleep depth (Fig.~\ref{multi}B). The emergence of a manifold in latent space where the sequence of correlation matrices was mapped preserving its continuity suggests that a low-dimensional representation can capture the signatures of progressively fading wakefulness.

The latent space could be divided into regions corresponding to wakefulness and all sleep stages, while also respecting the ordering of brain states in the descent to deep sleep (Fig.~\ref{multi}A, right). We applied the decoder exhaustively throughout the latent space, obtaining a pairwise correlation matrix for each $z_1$, $z_2$ pair (Fig.~\ref{multi}B). Next, we computed the structural similarity index (SSIM) to compare each matrix obtained from the latent space to the matrices corresponding to wakefulness, N1, N2 and N3 sleep. SSIM is defined as  $(\frac{2\mu_x\mu_y+0.01}{\mu_x^2+\mu_y^2+0.01})(\frac{2\sigma_x\sigma_y+0.03}{\sigma_x^2+\sigma_y^2+0.03})
(\frac{\sigma_{xy}+0.015}{\sigma_x\sigma_y+0.015})$,
where $x$ stands for each  $C_{ij}$ matrix shown in Fig.~\ref{multi}B and $y$ is the average $C_{ij}$ computed for each brain state.  The variables $\mu_x$,$\mu_x$,$\sigma_x$, $\sigma_y$ and $\sigma_{xy}$ correspond to the local means, standard deviations and co-variances of matrices $x$ and $y$ respectively. SSIM has the advantage of simultaneously weighting the Euclidean and correlation distances between matrices \cite{ipina2020modeling}. For each $z_1$, $z_2$ pair, we determined the brain state with the highest SSIM value and constructed the latent space parcellation shown in Fig.~\ref{multi}A (right panel). Again, we observe that the latent space can be orderly divided into regions corresponding to different sleep depth only from the model fitted to wakefulness and N3 sleep.

\begin{figure}[b]
\includegraphics[scale=0.8]{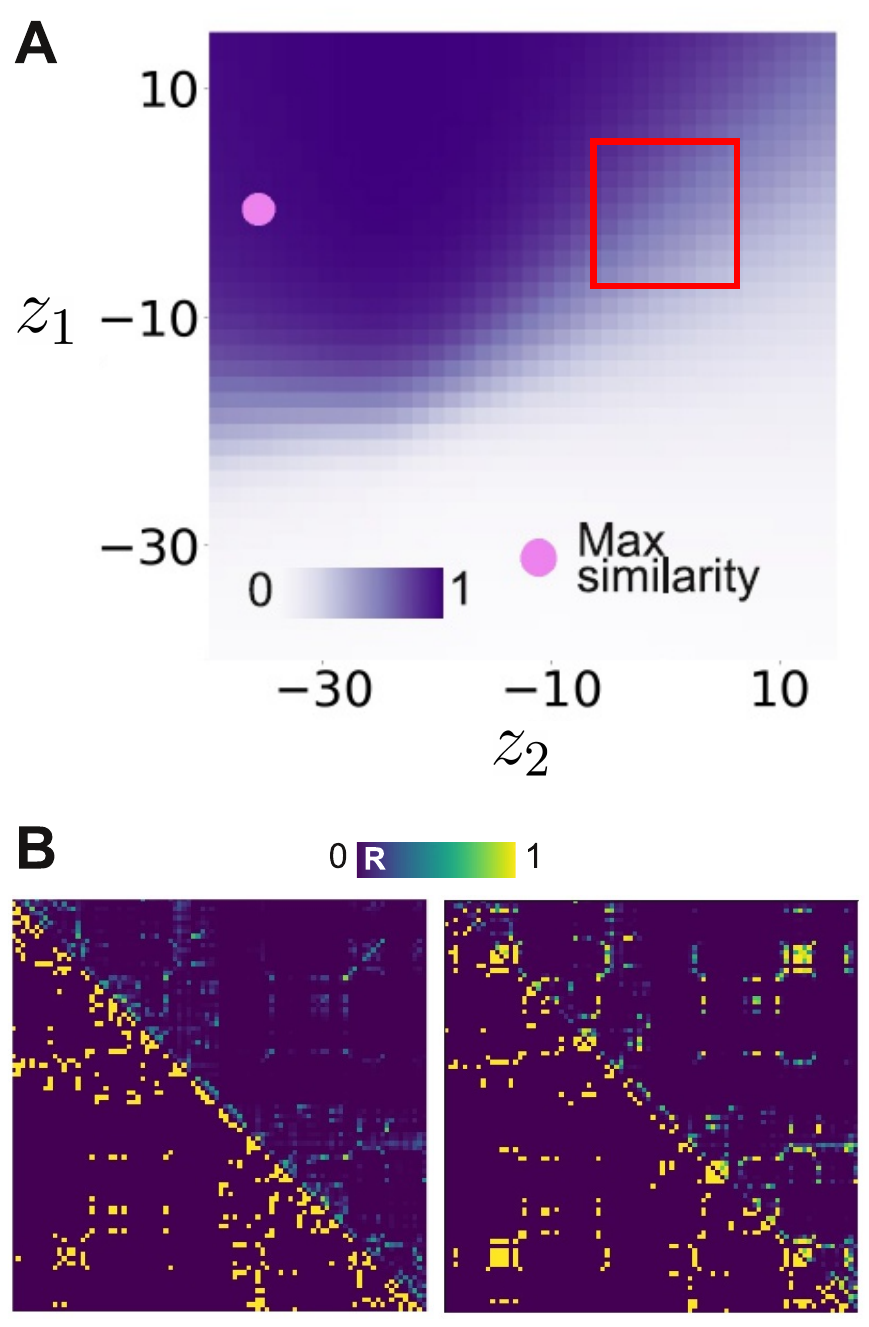}
\caption{\label{SC} Latent space variables can be extrapolated to reproduce increased structure-function coupling as a signature of unconsciousness. A) An exhaustive exploration of the SSIM between the decoded correlation matrices and the empirical structural connectivity matrix. High $z_1$ and low $z_2$ maximise this similarity. The red rectangle indicates the range of $z_1$ and $z_2$ reproduced in Fig.~\ref{multi}A. B) The empirical structural connectivity (left) and the best connectivity matrix reconstructed from the latent space (right) with the lower triangular part representing the matrices thresholded at 0.2.}
\end{figure}

\emph{Extreme latent space values predict collapse into structural connectivity.—} After mapping the progression of brain states during sleep into the latent space, we investigated whether the variables $z_1$, $z_2$ could be extrapolated to reproduce signatures of other unconscious states. We hypothesised that moving past N3 sleep in the latent space manifold where the progression of brain states is represented would increase the similarity between $C_{ij}^*$ (decoded correlation matrices computed from the dynamics) and $J_{ij}$ (structural coupling matrix). As previously shown both in humans and non-human primates\cite{vincent2007intrinsic, barttfeld2015signature, tagliazucchi2016deep}, states of deep unconsciousness are characterised by the collapse of functional coupling to the underlying anatomical connectivity structure. 

We decoded a wider range of latent space variables and computed the SSIM between the output correlations and the structural connectivity. As shown in Fig.~\ref{SC}A, moving beyond the N3 region in Fig.~\ref{SC}A (high $z_1$, low $z_2$) increased the similarity of the generated correlations with the structural connectivity. Exploring a wider region of the latent space, we found the highest similarity between empirical ($J_{ij}$) and reconstructed ($J_{ij}^*$) structural connectivity given by SSIM$(J_{ij},J_{ij}^*)=0.81$. Fig.~\ref{SC}A (left) shows the empirical $J_{ij}$ and Fig.~\ref{SC}A (right) shows the best connectivity matrix reconstructed from the latent space variables; in both cases the part below the diagonal corresponds to the matrices thresholded at 0.2. As hypothesised, moving past the N3 region in latent space reproduced a well-known signature of deep unconsciousness. This suggests that the latent space constructed from wakefulness and N3 sleep not only represented intermediate stages, but also captured a manifold where an ampler range of levels of consciousness can be represented.

\begin{figure}[t]
\includegraphics[scale=0.65]{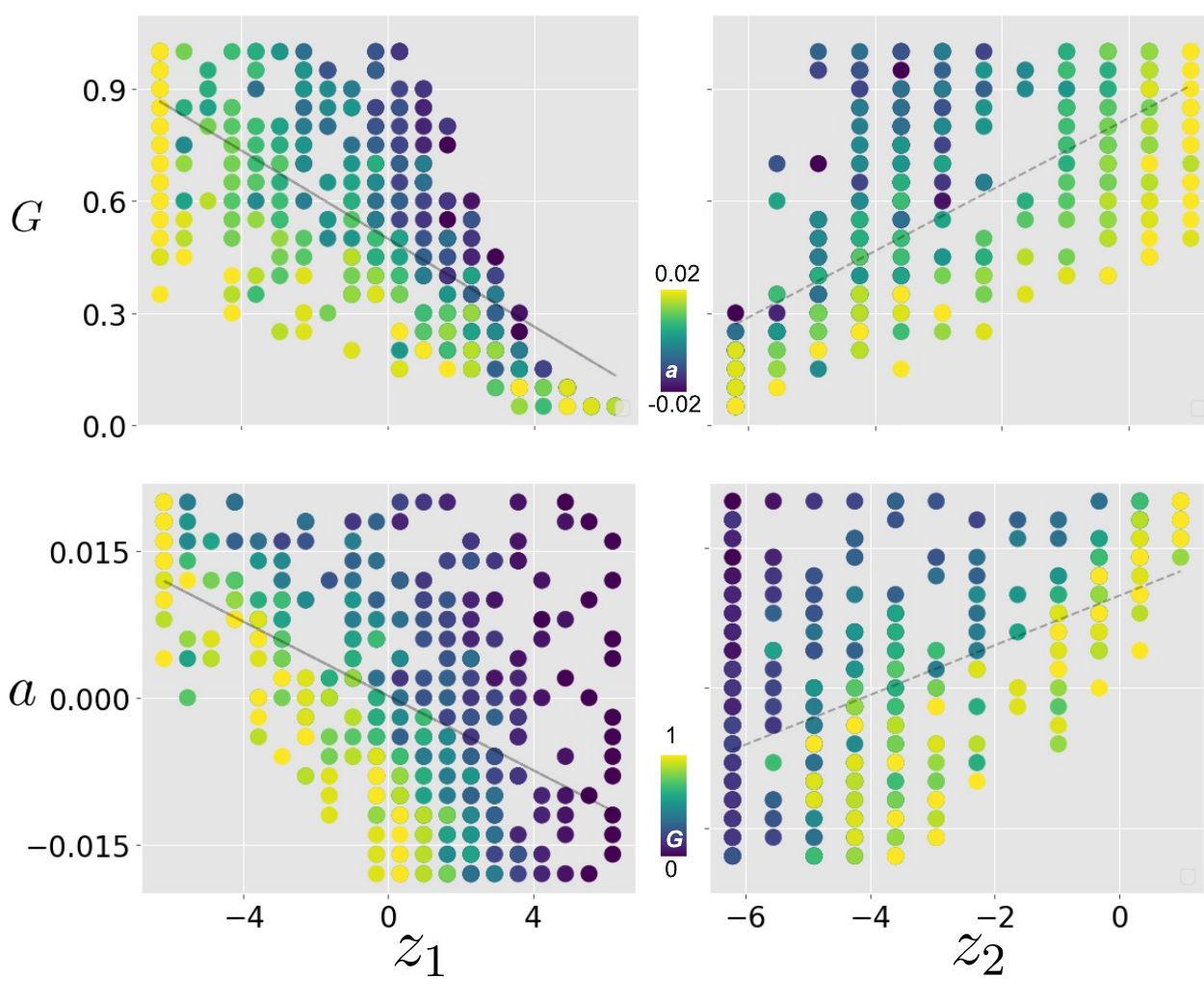}
\caption{\label{homo} Relationship between latent space variables ($z_1$, $z_2$) and the parameters of the homogeneous model ($a$, $G$).}
\end{figure}

\emph{Mapping the homogeneous model into the latent space.—} To gain further understanding concerning how the VAE successfully captured the progression of brain states from few parameter combinations, we trained a VAE using an homogeneous version of the nonlinear coupled oscillators in Eq. \ref{eq:one} (i.e. same $a$ for all oscillators), and compared the latent space encoding in variables $z_1$, $z_2$ with the parameters $a$ and $G$ \cite{jobst2017increased}. While the resulting correlation matrices do not reflect those obtained from the empirical data, the homogeneous model can be used to gain insight on the mapping performed by the VAE. 

We trained a VAE with $8000$ correlation matrices randomly extracted from a set of $1000$ matrices generated with the homogeneous model. Half of these matrices was generated using a high coupling factor ($G=0.8$) and a bifurcation parameter in the oscillatory regime ($a=0.015$) , while the other half was generated using low coupling ($G=0.2$) and a bifurcation parameter corresponding to fixed-point dynamics ($a=-0.015$). 

We decoded the latent space in 20 steps from -6.2 to 6.2 for each variable, obtaining a correlation matrix for each parameter combination. We also constructed several correlation matrices from the model with $a$ between -0.02 and 0.02 and $G$ between 0 and 1. For each parameter combination, we found the combination of latent space variables that maximised the SSIM between both matrices. In this way, we related each pair $(a,G)$ in the parameter space with each pair $(z_1,z_2)$ in the latent space. We found that both sets of variables were related by approximately linear relationships ($G$ vs. $z_1$, $r=-0.70$, $p<0.001$; $G$ vs.$z_2$, $r=0.69$, $p<0.001$; $a$ vs.$z_1$, $r=-0.56$, $p<0.001$; $a$ vs.$z_2$, $r=0.52$, $p<0.001$) (Fig.~\ref{homo}). This shows that for the simplified case of homogeneous $a$, the latent space approximates a linear transformation of the parameters governing the dynamics of the coupled oscillators. 

\emph{Discussion.—} Several recent studies demonstrated that low-dimensional models suffice to capture the large-scale correlation structure of neural activity seen during different brain states \cite{jobst2017increased, ipina2020modeling}. We went a step further, showing that these models implicitly represent different brain states as points in a low-dimensional manifold. This was established following a constructive process that consisted of training a VAE with correlation matrices belonging to a reduced set of brain states, and showing that the latent space represented intermediate states and could be extrapolated to produce hypothesised signatures of deeper unconsciousness. More generally, we showed that complex nonlinear dynamics depending on two parameters could be represented by a latent space that approximated a linear transformation of these parameters. Our results suggest that other (e.g. pathological) brain states could be captured and understood in terms of trajectories within a low-dimensional latent space, with potential applications in diagnosis, prognosis, and data augmentation. Generally, we propose that whenever complex collective dynamics are suspected to emerge from few independent parameters, VAE can be applied to reconstruct these parameters as a trajectory embedded in a low-dimensional latent space. 

Authors acknowledge funding from Agencia Nacional De Promocion Cientifica Y Tecnologica (Argentina), grant PICT-2018-03103.

\bibliographystyle{apsrev4-1} % Tell bibtex which bibliography style to use

\bibliography{apssamp}% Produces the bibliography via BibTeX.

\end{document}